\documentclass[12pt]{iopart}

\usepackage{amssymb}
\usepackage{setstack}
\usepackage[dvips]{graphicx}

\newcommand{\eqb}{\begin{equation}}
\newcommand{\eqe}{\end{equation}}
\newcommand{\eqbnon}{\begin{equation*}}
\newcommand{\eqenon}{\end{equation*}}

\newcommand{\eqab}{\begin{eqnarray}}
\newcommand{\eqae}{\end{eqnarray}}
\newcommand{\eqabnon}{\begin{eqnarray*}}
\newcommand{\eqaenon}{\end{eqnarray*}}

\newcommand{\defeq}{:=}
\newcommand{\defeqr}{=:}

\newcommand{\pd}[2]{\frac{\partial #1}{\partial #2}}

\newcommand{\eqref}[1]{\eref{#1}}

\newcommand{\vol}{\mbox{\boldmath $\varepsilon$}}
\newcommand{\pot}{\mbox{\boldmath $\Theta$}}
\newcommand{\binormal}{\mbox{\boldmath $\bar{\epsilon}$}}
\newcommand{\lie}{\mathcal{L}}

\newcommand{\Teff}{T_{\rm eff}}
\newcommand{\peff}{p_{\rm eff}}

\eqnobysec
\begin{document}

\title[First law of BH spacetime with $\Lambda$ and its application to SdS]{Mechanical First Law of Black Hole Spacetimes with Cosmological Constant and Its Application to Schwarzschild-de~Sitter Spacetime}

\author{Miho Urano$^1$ and Akira Tomimatsu$^2$}
\address{Department of Physics, Graduate School of Science, Nagoya University, Nagoya 464-8602, Japan}
\ead{$^1$\,urano@gravity.phys.nagoya-u.ac.jp}
\ead{$^2$\,atomi@gravity.phys.nagoya-u.ac.jp}

\author{Hiromi Saida}
\address{Department of Physics, Daido Institute of Technology, Nagoya 457-8530, Japan
~\footnote{Name of institution changes to Daido University at April 2009.}}
\ead{saida@daido-it.ac.jp}

\begin{abstract}
The mechanical first law (MFL) of black hole spacetimes is a geometrical relation which relates variations of mass parameter and horizon area. 
While it is well known that the MFL of asymptotic flat black hole is equivalent to its thermodynamical first law, however we do not know the detail of MFL of black hole spacetimes with cosmological constant which possess black hole and cosmological event horizons. 
Then this paper aims to formulate an MFL of the two-horizon spacetimes. 
For this purpose, we try to include the effects of two horizons in the MFL. 
To do so, we make use of the Iyer-Wald formalism and extend it to regard the mass parameter and the cosmological constant as two independent variables which make it possible to treat the two horizons on the same footing. 
Our extended Iyer-Wald formalism preserves the existence of conserved Noether current and its associated Noether charge, and gives the abstract form of MFL of black hole spacetimes with cosmological constant. 
Then, as a representative application of that formalism, we derive the MFL of Schwarzschild-de~Sitter (SdS) spacetime. 
Our MFL of SdS spacetime relates the variations of three quantities; the mass parameter, the total area of two horizons and the volume enclosed by two horizons. 
If our MFL is regarded as a thermodynamical first law of SdS spacetime, it offers a thermodynamically consistent description of SdS black hole evaporation process: The mass decreases while the volume and the entropy increase. 
In our suggestion, the generalized second law is not needed to ensure the second law of SdS thermodynamics for its evaporation process.
\end{abstract}
\submitto{Classical and Quantum Gravity \, \rm (Accepted for Publication, 23 March 2009)}
\maketitle

\section{Introduction}
\label{sec:intro}

Black hole thermodynamics has already been established well for spacetimes with a single black hole event horizon~\cite{ref:bht}. 
However there is no consistent thermodynamical formulation for black hole spacetimes with cosmological constant $\Lambda$ which possess a black hole event horizon (BEH) and a cosmological event horizon (CEH).
For example, in Schwarzschild-de~Sitter (SdS) spacetime, the Hawking temperature of BEH is higher than that of CEH~\cite{ref:sds.temperatures}. 
This temperature difference causes difficulties in formulating SdS thermodynamics, since BEH (CEH) seems in some non-equilibrium state under the influence of Hawking radiation of different temperature by CEH (BEH). 
In general, the similar difficulty arises also for any two-horizon spacetime with $\Lambda$.
\footnote{
Some proposals for thermodynamics of BEH and CEH are already given for a case with some special matter fields and for an extreme case with magnetic/electric charge~\cite{ref:sds.matter}. 
These examples are artificial to vanish the temperature difference of BEH and CEH. 
But in this paper we consider more general case which is not extremal and does not depend on artificial matter fields.
}

However we may be able to search for thermodynamical first law of two-horizon spacetimes with $\Lambda$ through the \emph{mechanical first law} (MFL). 
The MFL is a geometrical relation which relates the variations of mass parameter, horizon area and the other supplemental quantities. 
For example, the MFL of Schwarzschild black hole is equivalent to its thermodynamical first law, and formulated by using the mass parameter as only one variable which describes the effect of a single horizon on the MFL. 
This fact of Schwarzschild black hole leads us to expect that the MFL of two-horizon spacetimes with $\Lambda$ can be a candidate of a thermodynamical first law of the spacetimes, and that the existence of two horizons requires two independent variables in the resultant MFL of the spacetimes to include the effects of two horizons. 
Then we adopt the following working hypothesis to search for the MFL of two-horizon spacetimes with $\Lambda$:
\begin{description}
\item[Working Hypothesis (Two Independent Variables):]
Generally for spacetimes with cosmological constant which possess BEH and CEH, the mass parameter $M$ and the cosmological constant $\Lambda$ are regarded as two independent variables in the MFL of the spacetimes.
\end{description}
Indeed it will be shown in Sec.\ref{sec:mfl} that the MFL of SdS spacetime becomes mathematically inconsistent if~$\Lambda$ is not an independent variable. 
When one considers a non-variable $\Lambda$ as a physical situation, it is obtained by setting the variation of~$\Lambda$ zero ($\delta \Lambda = 0$) in the MFL after constructing it with regarding~$\Lambda$ as an independent variable. 
In such case, the variable~$\Lambda$ is interpreted as a ``working variable'' to obtain the MFL of two-horizon spacetimes with~$\Lambda$.

On the other hand for SdS spacetime, the search for an MFL with and without regarding $\Lambda$ as an independent variable, has already been tried in some existing works~\cite{ref:sds.mfl,ref:sds.mfl_2}. 
Those works make use of some conserved quantities defined by some integrals (e.g.~Hamiltonian, action integral and so on). 
One can obtain some candidates of MFL or thermodynamical first law of SdS spacetime with regarding the integral quantities as the variables in the first law (see~\cite{ref:sds.mfl_2} for example).
However, when one uses the integral quantities, there arises a problem of the choice of integration constant (e.g. the so-called \emph{boundary counter-term} or \emph{subtraction term} to eliminate some divergent term in Hamiltonian or action integral).

Then, in order to derive the MFL of two-horizon spacetimes with $\Lambda$, we make use of the Iyer-Wald formalism which is free from the problem of integration constant~\cite{ref:iw}. 
We extend the original Iyer-Wald formalism to make it so general to be applicable to any spacetime with $\Lambda$ under the working hypothesis of two independent variables. 
The extended Iyer-Wald formalism preserves the existence of conserved Noether current and its associated Noether charge, and enables us to observe how the resultant MFL is related to the Noether charge which is locally constructed from the metric (free from the problem of integration constant).

The general extension of Iyer-Wald formalism to include the variable $\Lambda$ gives simply an abstract form of MFL of two-horizon spacetimes with $\Lambda$, since the metric is not concretely specified. 
Then, as a representative application of our extended Iyer-Wald formalism, we derive the MFL of SdS spacetime by that formalism. 
Moreover, we examine what the MFL of SdS spacetime implies for SdS black hole evaporation if the MFL is regarded as a thermodynamical first law.

Here let us note that Ref.\cite{ref:sds.mfl_2} has already treated $\Lambda$ as an independent variable for SdS spacetime. 
(In~\cite{ref:sds.mfl}, $\Lambda$ is not an independent variable but a complete constant.) 
In~\cite{ref:sds.mfl_2}, the variable $\Lambda$ is concluded from mathematical consistency of the MFL of SdS spacetime, although there remains an uncertainty of integration constant in the resultant MFL given in~\cite{ref:sds.mfl_2}. 
However we will show in this paper that the extended Iyer-Wald formalism gives the same MFL with Ref.\cite{ref:sds.mfl_2} in a straightforward way without the problem of integration constant. 
Furthermore, as the advantage of our procedure, the extended Iyer-Wald formalism enables us to introduce naturally a state variable which represents the size of the two-horizon system. 
Then, consequently, we will show that the new state variable of system size leads natural definitions of effective temperature and pressure of two-horizon system. 
Those state variables are suitable to describe the SdS black hole evaporation process. 
Such good state variables are not given in~\cite{ref:sds.mfl_2} and also the evaporation process is not described in~\cite{ref:sds.mfl_2}.

This paper is organized as follows. 
Sec.\ref{sec:iw} is devoted to the extension of Iyer-Wald formalism to include the variable~$\Lambda$ and obtain the abstract form of MFL of any two-horizon spacetimes with $\Lambda$. 
Sec.\ref{sec:mfl} is for the application of extended Iyer-Wald formalism to SdS spacetime and obtain the MFL of SdS spacetime. 
In that section, the state variables of system size, effective temperature and pressure are given, and then the mathematical consistency of the MFL with variable $\Lambda$ is also examined. 
In Sec.\ref{sec:d}, we discuss what is implied about SdS black hole evaporation if our MFL of SdS spacetime is regarded as a thermodynamical first law, and give a comment about the non-equilibrium nature due to the difference of Hawking temperatures of BEH and CEH.

Throughout this paper, we use the Planck units, $c = \hbar = G = k_B =1$.

\section{Iyer-Wald formalism with variable $\Lambda$}
\label{sec:iw}

For simplicity we consider the empty spacetime in $n$ dimensions without matter fields and start with the diffeomorphism invariant Lagrangian $n$-form ${\bf L}(g,\Lambda) = L(g,\Lambda)\,\vol$, where $g$ denotes the metric $g_{\mu \nu}$, $\vol$ is the volume $n$-form and $L(g,\Lambda)$ is the Lagrangian scalar density. 
The inclusion of matter fields is very straightforward in following discussions. 
In the next section, ${\bf L}$ is specified to be the ordinary Einstein-Hilbert form in four dimensions, 
\eqb
\label{eq:iw.L}
 {\bf L} = \frac{1}{16\pi}\left( \mathcal{R} - 2\Lambda \right) \, \vol \, ,
\eqe
where $\mathcal{R}$ is the Ricci scalar and $\Lambda$ is the variable cosmological constant.
But in this section ${\bf L}(g,\Lambda)$ is an arbitrary Lagrangian of the metric and the variable~$\Lambda$. 
The first variation of this Lagrangian is expressed as
\eqb
\label{eq:iw.delta_L}
 \delta {\bf L} =  {\bf E}_g \delta g
                 + d\pot( g , \delta g )
                 + {\bf E}_{\Lambda}  \delta\Lambda \, ,
\eqe
where $\delta g$ is the abbreviation of metric variation $\delta g_{\mu\nu}$, $\pot$ is the ($n-1$)-form called the \emph{symplectic potential} which corresponds to the boundary term in the variational principle of action integral, ${\bf E}_g = E_g\,\vol$ gives the Einstein equation of the metric by $E_g = 0$ and ${\bf E}_{\Lambda} = E_{\Lambda}\,\vol$ is the variation of ${\bf L}$ with respect to the working variable $\Lambda$,
\eqb
 {\bf E}_{\Lambda} = \pd{L}{\Lambda}\,\vol \, .
\eqe
For ${\bf L}$ in Eq.\eqref{eq:iw.L}, we get ${\bf E}_{\Lambda} = -(1/8\pi)\,\vol$ which will be used in the next section.

Here we have to emphasize the following two remarks: 
The first one is on the so-called \emph{on-shell condition}, which restricts the metric (and the matter field if it exists) to be the solution of equation of motion. 
When the on-shell condition in the presence of metric variation is going to be required later in this section, the variation $\delta g_{\mu\nu}$ is to be understood as a solution of the ``extended'' linearized Einstein equation due to the variable $\Lambda$,
\eqb
\label{eq:iw.linear}
\delta G_{\mu\nu} + \Lambda\,\delta g_{\mu\nu} + g_{\mu\nu}\,\delta\Lambda = 0 \,,
\eqe
where $G_{\mu\nu}$ is the Einstein tensor and $g_{\mu\nu}$ is the ``unperturbed'' metric given by the Einstein equation $E_g = 0$~\footnote{This $g_{\mu\nu}$ can also be regarded as a ``background'' metric of the perturbation $\delta g_{\mu\nu}$.}. 
(The right-hand side would not be zero but be the variation $\delta T_{\mu\nu}$ of stress-energy tensor if matter fields exit.) 
The third term $g_{\mu\nu}\,\delta\Lambda$ does not appear for the ordinary linearized equation with completely constant $\Lambda$, but now it appears due to the variable~$\Lambda$. 
Under the working hypothesis of two independent variables, we can regard~$\Lambda$ not as the kin of universal constants like Newton's constant but as the kin of constants of motion like mass parameter~$M$ which is a variable in asymptotic flat black hole thermodynamics.

The second remark we should emphasize here is on the variational principle. 
Exactly speaking, the variational principle gives the equations of motion of dynamical variables via the vanishing variation of action integral $I$ with respect to the dynamical variable $\phi(x)$ which depends generally on the spacetime points, $\delta I/\delta\phi(x) = 0$, where $x$ represents the spacetime dependence. 
However, the ``working'' variable $\Lambda$ is not regarded as any dynamical variable and has no spacetime dependence. 
The $\Lambda$ is simply a working variable to ensure the mathematical consistency of the resultant MFL of two-horizon spacetimes. 
Hence, even if ${\bf E}_{\Lambda}$ is set zero, it can never be interpreted as any equation of motion of dynamical variable. 
This means that the on-shell condition requires only the Einstein equation $E_g = 0$ (and the extended linearized equation~\eref{eq:iw.linear} under the presence of variation $\delta g_{\mu\nu}$), and ${\bf E}_{\Lambda}$ is non-vanishing (${\bf E}_{\Lambda} \not\equiv 0$). 
Now it is recognized that the physical principle we rely on is the ``extended'' variational principle in which the variations are taken with respect to not only the dynamical variable $g_{\mu\nu}$ (and matter field if it exists) but also the working variable $\Lambda$, while the equations of motion are given by the variation of Lagrangian with respect to dynamical variables.

As a by-product of the above two remarks, it will be shown below that, in our extended Iyer-Wald formalism, the \emph{conserved Noether current} and its \emph{associated Noether charge} are defined in the same way as in the original Iyer-Wald formalism. 
Our ``extension'' of Iyer-Wald formalism has three meanings; (1)~the extended Iyer-Wald formalism includes the variable~$\Lambda$, (2)~the $\Lambda$ is not a dynamical variable but simply the working variable which means ${\bf E}_{\Lambda}$ does not give any equation of motion, and (3)~the conserved Noether current is obtained with the same definition as in the original Iyer-Wald formalism.

Then let us proceed to the extension of Iyer-Wald formalism. 
Eq.\eqref{eq:iw.delta_L} is the starting point. 
The extended Iyer-Wald formalism with variable $\Lambda$ differs from the original formalism on the following two points: 
One of them is a manifest point expressed by the third term in the right-hand side of Eq.\eqref{eq:iw.delta_L}. 
That term does not arise in the original Iyer-Wald formalism, but arises in our extended formalism by the variation $\delta\Lambda$ in, for example, the second term in the right-hand side of Eq.\eqref{eq:iw.L}. 
Another point is a subtle point included in the metric variation $\delta g_{\mu\nu}$. 
In our extended formalism, the variation $\delta g_{\mu\nu}$ also gives rise to the variation $\delta\Lambda$ if the concrete form of the metric depends on $\Lambda$ as for SdS, de~Sitter and Anti-de~Sitter spacetimes.

To formulate the abstract form of MFL with the variable $\Lambda$, let us follow the same procedure of the original Iyer-Wald formalism~\cite{ref:iw}. 
If we introduce an arbitrary vector field $\xi^{\mu}$, which is not a dynamical variable in ${\bf L}$, and consider the variation given by the Lie derivative $\delta = \lie_{\xi}$ along $\xi^{\mu}$, then we get from Eq.\eqref{eq:iw.delta_L},
\eqb
\label{Lagrangean_lambda_noethercurrent}
 d\left[ \pot(g , \lie_\xi g) - \xi \cdot {\bf L} \right] =
   - {\bf E}_g \lie_\xi g - {\bf E}_{\Lambda} \lie_\xi \Lambda \, ,
\eqe
where $\xi\cdot{\bf L} \defeq \xi^{\mu} {\bf L}_{\mu \nu_1 \cdots \nu_{n-1}}$ and a relation $\lie_\xi {\bf A} = \xi\cdot d{\bf A} + d(\xi\cdot{\bf A})$ of the Lie and exterior derivatives of a form ${\bf A}$ is used. 
The ($n-1$)-from in the left-hand side, $\pot(g , \lie_\xi g) - \xi \cdot {\bf L} \defeqr {\bf J}_\xi$, is called the \emph{Noether current}. 
Here note that, since $\Lambda$ has no spacetime dependence, $\lie_\xi \Lambda \equiv 0$ holds. 
Therefore, when the on-shell condition $E_g = 0$ is required, the Noether current is closed $d{\bf J}_\xi = 0$ which guarantees the local existence of the ($n-2$)-form ${\bf Q}_\xi$ called the \emph{Noether charge},
\eqb
\label{eq:iw.Q}
 d{\bf Q}_\xi \defeq {\bf J}_\xi = \pot(g , \lie_\xi g) - \xi \cdot {\bf L} \, .
\eqe
This ${\bf Q}_\xi$ is locally constructed from the on-shell metric which satisfies the Einstein equation. 
Note that the conservation of the Noether current $d{\bf J}_\xi = 0$ under the on-shell condition corresponds to the \emph{Noether's theorem}, and the Noether charge ${\bf Q}_\xi$ is the conserved charge of ${\bf J}_\xi$ associated with the symmetry generator $\xi$. 
The definition of ${\bf Q}_\xi$ in Eq.\eqref{eq:iw.Q} is the same with that in the original Iyer-Wald formalism. 
Here it should also be emphasized that the existence condition of ${\bf Q}_\xi$ is the on-shell condition $E_g = 0$ for the metric $g_{\mu\nu}$ appearing in Eq.\eref{eq:iw.Q}. 
This condition is the same with that in the original Iyer-Wald formalism.
Hence we find that, in our extended Iyer-Wald formalism, the conserved Noether current and its associated Noether charge are defined in the same way as in the original Iyer-Wald formalism.

Next, since the vector $\xi^{\mu}$ (not the 1-form $\xi_{\mu}$) is not a dynamical variable, its variation does not exist ($\delta\xi^{\mu} \equiv 0$) under the variation of dynamical and working variables. 
Then, from the variation of ${\bf J}_\xi$, the following relation is obtained:
\eqab
\label{eq:iw.deltaJ}
\delta {\bf J}_\xi
 &=& \delta\pot(g ,\lie_\xi g) - \xi\cdot\delta {\bf L} \nonumber \\
 &=& {\bf \omega }(g ,\delta g ,\lie_\xi g) + d\left[ \xi\cdot\pot(g ,\delta g) \right]
        - \xi\cdot{\bf E}_\Lambda\,\delta\Lambda \, ,
\eqae
where ${\bf \omega}$ is defined as ${\bf \omega }(g ,\delta g ,\lie_\xi g) \defeq \delta\pot(g ,\lie_\xi g) - \lie_\xi \pot(g ,\lie_\xi g)$ and the on-shell condition $E_g = 0$ is required. 
Here we must note that, as explained in the second paragraph of this section, the on-shell condition under the presence of metric variation $\delta g_{\mu\nu}$ denotes that the unperturbed matric $g_{\mu\nu}$ and the variation $\delta g_{\mu\nu}$ appearing in Eq.\eref{eq:iw.deltaJ} satisfy, respectively, the Einstein equation $E_g = 0$ and the extended linearized Einstein equation~\eref{eq:iw.linear}. 
Furthermore, as noted in~\cite{ref:iw}, the $(n-1)$-form ${\bf \omega}$ vanishes (${\bf \omega} = 0$), when $\xi$ is the generator of a symmetry of all dynamical variables in ${\bf L}$, i.e. $\lie_\xi g_{\mu\nu} = 0$. 
For stationary spacetimes, the timelike Killing vector can be regarded as the symmetry generator $\xi$. 
Hence, at least for the stationary case, we get
\eqb
\label{eq:iw.delta_J}
 \xi\cdot{\bf E}_\Lambda\,\delta\Lambda +
 \delta {\bf J}_\xi - d\left[\xi\cdot\pot (g ,\delta g) \right] = 0 \, .
\eqe
Here the Noether charge defined in Eq.\eqref{eq:iw.Q} gives $\delta{\bf J}_\xi = d(\delta{\bf Q}_\xi)$. 
Therefore, by integrating Eq.\eqref{eq:iw.delta_J} on a hypersurface~$\Sigma$ and applying the Stokes' theorem, we obtain
\eqb
\label{eq:iw.mfl_primitive}
 \int_{\Sigma} \xi\cdot{\bf E}_{\Lambda}\delta\Lambda
 + \int_{\partial\Sigma}
   \left[\, \delta{\bf Q}_{\xi} - \xi\cdot\pot(g , \delta g) \,\right] = 0 \, .
\eqe
The first term is due to the variable $\Lambda$. 
There are three conditions to hold Eq.\eqref{eq:iw.mfl_primitive}; (1)~the Einstein equation $E_g = 0$ for the unperturbed metric $g_{\mu\nu}$, (2)~the extended linearized Einstein equation~\eref{eq:iw.linear} of the metric variation $\delta g_{\mu\nu}$, and (3)~the existence of the symmetry generator~$\xi$ to give $\lie_\xi g = 0$.

In the original Iyer-Wald formalism, the first term in Eq.\eqref{eq:iw.mfl_primitive} disappears. 
Ref.\cite{ref:iw} shows that, when the original formalism is applied to asymptotic flat black holes, Eq.\eqref{eq:iw.mfl_primitive} without the first term reduces to the MFL of those black holes. 
Therefore we can regard Eq.\eqref{eq:iw.mfl_primitive} as the primitive MFL of any spacetime with $\Lambda$ in any dimensions.~\footnote{
The primitive MFL~\eqref{eq:iw.mfl_primitive} seems applicable to spacetimes without CEH like asymptotic Anti-de~Sitter black holes. 
When Eq.\eqref{eq:iw.mfl_primitive} is applied to such black holes, the two degrees of freedom expressed by $M$ and $\Lambda$ may describe two independent effects in the MFL; one of them is due to the BEH, and another is due to the ``wall'' given by the infinitely large gravitational potential barrier of Anti-de~Sitter metric.
}

\section{Mechanical first law of SdS spacetime}
\label{sec:mfl}

\subsection{Preparations}

Before deriving the MFL of SdS spacetime from Eq.\eqref{eq:iw.mfl_primitive}, we summarize the SdS metric and the MFL of asymptotic flat black holes derived from the original Iyer-Wald formalism. 
For the first, the line element of SdS spacetime is
\eqb
\label{eq:mfl.metric}
 ds^2 = - f(r)\,dt^2  + \frac{dr^2}{f(r)}
       + r^2\,\left( d\theta^2  + \sin^2 \theta d\varphi^2 \right) \, ,
\eqe
where 
\eqb
\label{eq:mfl.f}
 f(r) \defeq 1 - \frac{2M}{r} - \frac{\Lambda}{3}\,r^2 \, .
\eqe
The equation $f(r) = 0$ has two positive roots and one negative root for the parameter range,
\eqb
\label{eq:mfl.parameter_range}
 0 < 9\,M^2\,\Lambda < 1 \, .
\eqe
Throughout this paper, our discussion is restricted in this range. 
This parameter range guarantees the existence of BEH and CEH. 
The radius of BEH $r_b$ and that of CEH $r_c$ are given by, respectively, the smaller positive root and the larger positive root of $f(r)=0$. 
Then the parameter range~\eqref{eq:mfl.parameter_range} gives the relations of these radii,
\eqb
\label{eq:mfl.radii}
 2M < r_b  < 3M < \frac{1}{\sqrt{\Lambda}} < r_c  < \frac{3}{\sqrt{\Lambda}} \, .
\eqe
The equations $f(r_b) = 0$ and $f(r_c) = 0$ are rearranged to
\eqb
\label{eq:mfl.M.Lambda}
 M = \frac{r_b\,r_c\,\left( r_b + r_c \right)}
             {2\,\left( r_b^2 + r_b \, r_c + r_c^2 \right)} \quad, \quad
 \Lambda = \frac{3}{r_b^2  + r_b \, r_c  + r_c^2} \, .
\eqe
Since SdS spacetime is static, each event horizon possesses the bifurcation sphere, on which the timelike Killing vector $\xi \defeq \partial_t$ vanishes. 
The surface gravity of BEH $\kappa_b$ and that of CEH $\kappa_c$ are defined by the following equations evaluated at the horizons,
\eqb
\label{eq:mfl.kappa_def}
 \xi^{\nu}\nabla_{\nu} \xi^{\mu} \Bigr|_{\rm BEH} = \kappa_b \, \xi^{\mu} \Bigr|_{\rm BEH}
 \quad,\quad
 \xi^{\nu}\nabla_{\nu} \xi^{\mu} \Bigr|_{\rm CEH} = \kappa_c \, \xi^{\mu} \Bigr|_{\rm CEH} \,.
\eqe
In general, the numerical value of surface gravity depends on the normalization of $\xi$. 
With the normalization $\xi \defeq \partial_t$, we get
\eqab
\label{eq:mfl.kappa}
 \kappa_b = \left.\frac{1}{2}\frac{d f(r)}{dr}\right|_{r=r_b}
   = \frac{1 - \Lambda\,r_b^2}{2\,r_b} \quad,\quad
 \kappa_c &=& \left.\frac{1}{2}\frac{d f(r)}{dr}\right|_{r=r_c}
   = \frac{1 - \Lambda\,r_c^2}{2\,r_c}  \, .
\eqae
Then, from Eq.\eqref{eq:mfl.radii}, the following relations hold,
\eqb
\label{eq:mfl.kappa_relations}
 \kappa_b > 0 \quad,\quad \kappa_c < 0 \quad,\quad
 \kappa_b > \left|\kappa_c\right| \, .
\eqe

Next, we summarize the MFL of asymptotic flat black holes derived from the original Iyer-Wald formalism~\cite{ref:iw}: 
The first term in Eq.\eqref{eq:iw.mfl_primitive} does not exist in the original formalism. 
When the original formalism is applied to asymptotic flat black holes and $\Sigma$ connects the bifurcation sphere $B$ of the horizon and the spatial infinity~$\infty$, then Eq.\eqref{eq:iw.mfl_primitive} without the first term is rearranged to
\eqb
\label{eq:mfl.mfl_flat_primitive}
 \int_{\infty}
 \left[\, \delta{\bf Q}_{\xi} - \xi\cdot\pot(g , \delta g) \,\right] -
 \int_B \delta{\bf Q}_\xi = 0 \, ,
\eqe
where $\xi \defeq \partial_t$ with the usual static time $t$ and $\xi = 0$ on $B$ is used. 
As shown in~\cite{ref:iw}, this reduces to the MFL of asymptotic flat black holes,
\eqb
\label{eq:mfl.mfl_flat}
  \delta M = \frac{\kappa_{\rm BH}}{8\pi}\,\delta A_{\rm BH} + \mbox{work terms} \, ,
\eqe
where $M$ is the ADM mass of black hole, $A_{\rm BH}$ is the area of $B$, $\kappa_{\rm BH}$ is the surface gravity of the horizon with an appropriate normalization of Killing vector, the ``work terms'' are given by electromagnetic fields and angular momentum for Reissner-Nordstr\"{o}m and Kerr black holes~\cite{ref:4laws}. 
(This MFL in Eq.\eqref{eq:mfl.mfl_flat} is a differential form of the so-called \emph{Smarr's formula}~\cite{ref:mf.smarr}.) 
It should be emphasized that the first term (integration on $\infty$) in the left-hand side of Eq.\eqref{eq:mfl.mfl_flat_primitive} gives the variation of the ADM mass $\delta M$ and the ``work terms'' in Eq.\eqref{eq:mfl.mfl_flat}, and the second term (integration on $B$) in Eq.\eqref{eq:mfl.mfl_flat_primitive} gives the ``heat term'' $(\kappa_{\rm BH}/8\pi)\,\delta A_{\rm BH}$. 
In asymptotic flat black hole thermodynamics, the MFL in Eq.\eqref{eq:mfl.mfl_flat} is equivalent to the thermodynamical first law of asymptotic flat black holes in which $M$ is regarded as the internal energy, and $A_{\rm BH}/4$ is regarded as the entropy (the so-called \emph{entropy-area law}) due to the Hawking temperature $\kappa_{\rm BH}/2 \pi$~\cite{ref:bht}.

\subsection{Application of extended Iyer-Wald formalism to SdS spacetime}

Now we apply Eq.\eqref{eq:iw.mfl_primitive} to SdS spacetime to obtain the \emph{appropriately expressed} MFL of SdS spacetime. 
Here the meaning of ``appropriate expression'' is that the mass parameter $M$ of SdS spacetime is expressed as a function of two independent quantities which seem correspond to the entropy and the state variable of system size when the MFL is regarded as a thermodynamical first law. 
The quantity corresponding to entropy gives the ``heat term'' in MFL, and the quantity corresponding to system size gives the ``work term'' in MFL. 
In thermodynamical first law of ordinary laboratory systems, the heat term is the product of temperature and variation of entropy, and the work term is the product of pressure and variation of system size.

Concerning the quantity $S$ which seems correspond to the entropy of SdS spacetime, we refer simply to the entropy-area law of asymptotic flat black holes and define $S$ by the total spatial area of BEH and CEH: 
\eqb
\label{eq:mfl.S}
 S \defeq \pi\,r_c^2 + \pi\,r_b^2 \,.
\eqe
Here note that the horizon radii $r_b$ and $r_c$ are regarded as two independent variables due to the working hypothesis of two independent variables, and that the total horizon area includes manifestly the effects of BEH and CEH through these two variables $r_b$ and $r_c$. 
When this definition~\eqref{eq:mfl.S} is adopted, the variation $\delta S$ gives the ``heat term'' in the appropriate MFL of SdS spacetime.

Next, we have to define the system size $V$ whose variation $\delta V$ composes the ``work term'' in the appropriate MFL of SdS spacetime. 
Here, for the system size of SdS spacetime, it seems reasonable to consider the three dimensional volume of a spacelike hypersurface connecting BEH and CEH as the object of thermodynamical interest, since the Hawking radiations of two horizons coexist there. 
As such hypersurface, let us consider $\Sigma_{\rm bif}$ which connects the bifurcation spheres of BEH and CEH at $t =$ constant. 
Furthermore let us refer to the first term in Eq.\eqref{eq:iw.mfl_primitive}. 
In that term, the integral $\int_{\Sigma_{\rm bif}} \xi\cdot {\bf E}_\Lambda$ appears as a natural quantity of three dimensional volume. 
Then concerning the quantity $V$ which seems correspond to the system size of SdS spacetime, we adopt the following definition:
\eqb
\label{eq:mfl.V}
 V \defeq \int_{\Sigma_{\rm bif}} \xi\cdot {\bf E}_\Lambda = 
          \frac{4\,\pi}{3} \left( r_c^3 - r_b^3 \right) \, ,
\eqe
where $\xi \defeq \partial_t$.

The definitions~\eqref{eq:mfl.S} and~\eqref{eq:mfl.V} are equivalent to search for the MFL of SdS spacetime which is expressed as
\eqb
\label{eq:mfl.mfl_def}
 \delta M = \Teff\,\delta S - \peff\,\delta V \, ,
\eqe
where the concrete forms of coefficients $\Teff$ and $\peff$ will be obtained below. 
These coefficients $\Teff$ and $\peff$ are simply the partial derivatives of $M$. 
If the MFL in Eq.\eqref{eq:mfl.mfl_def} is regarded as a thermodynamical first law of SdS spacetime, then $\Teff$ and $\peff$ are interpreted as, respectively, effective temperature and pressure. 
Therefore, because the temperature and pressure are positive definite for ordinary laboratory systems, it is natural to require that the appropriately expressed MFL should satisfy the following requirement:
\eqb
\label{eq:mfl.req}
 \Teff > 0 \quad,\quad \peff > 0 \, .
\eqe
If the MFL in Eq.\eqref{eq:mfl.mfl_def} is regarded as a thermodynamical first law of SdS spacetime, then the positivity of $\Teff$ and $\peff$ makes it plausible to call $\Teff$ and $\peff$, respectively, the effective temperature and pressure.
A comment on $\Teff$ in relation to non-equilibrium nature of SdS spacetime will be given in Sec.\ref{sec:d}.

The appropriately expressed MFL of SdS spacetime should be obtained by using the definitions~\eqref{eq:mfl.S} and~\eqref{eq:mfl.V} under the requirement~\eqref{eq:mfl.req} with regarding $M$ and $\Lambda$ as two independent variables. 
To obtain the appropriate MFL, we must be careful to carry out the integrations in Eq.\eqref{eq:iw.mfl_primitive}. 
According to asymptotic flat case summarized in the previous subsection, we expect that, when the second term (surface term) in Eq.\eqref{eq:iw.mfl_primitive} is evaluated on the bifurcation sphere in SdS spacetime, it should give the ``heat term'' $\Teff\,\delta S$ in the MFL of SdS spacetime~\eqref{eq:mfl.mfl_def}. 
But, when the second term in Eq.\eqref{eq:iw.mfl_primitive} is evaluated on some sphere which is not the bifurcation sphere, it should contribute to the mass variation $\delta M$ in Eq.\eqref{eq:mfl.mfl_def}, as for the ADM mass in asymptotic flat case~\eqref{eq:mfl.mfl_flat}. 
And the first term (3-volume term) in Eq.\eqref{eq:iw.mfl_primitive} should contribute to the ``work term'' $-\peff\,\delta V$ in Eq.\eqref{eq:mfl.mfl_def}. 
Then, since there seems no natural sphere except for the bifurcation spheres of horizons, one problem arises; how to prepare the sphere to get the mass variation in the MFL of SdS spacetime. 
Hence, in order to extract the mass variation from Eq.\eqref{eq:iw.mfl_primitive}, we adopt the strategy composed of three steps as follows:
\begin{description}
\item[Step 1:]
Consider two hypersurfaces, $\Sigma_b$ and $\Sigma_c$, at $t =$~constant. 
Here $\Sigma_b$ covers the region $r_b < r < r_{cut}$ and $\Sigma_c$ covers $r_{cut} < r < r_c$, where $r_{cut}$ is a working variable which is introduced to extract the mass variation $\delta M$ and must not appear in the resultant MFL of SdS spacetime. 
\item[Step 2:]
Carry out the integrations in Eq.\eqref{eq:iw.mfl_primitive} for the hypersurface $\Sigma_b$. 
Do the same for~$\Sigma_c$.
\item[Step 3:]
Combine those integrated equations to introduce the variations of $M$, $S$ and $V$ into the appropriately expressed MFL of SdS spacetime~\eqref{eq:mfl.mfl_def}. 
\end{description}
Note that the working variable $r_{cut}$ is introduced simply to carry out the integrations in Eq.\eqref{eq:iw.mfl_primitive} as a ``piecewise'' integration. 
If the resultant MFL obtained in the step~3 depends on $r_{cut}$, this strategy is not suitable for the construction of the appropriate MFL of SdS spacetime. 
However in fact, as will be shown below, the result does not depend on $r_{cut}$ and we will succeed to obtain the appropriately expressed MFL of SdS spacetime.

Let us carry out the step~2. 
We calculate Eq.\eqref{eq:iw.mfl_primitive} with the hypersurfaces~$\Sigma_b$ and~$\Sigma_c$ which are introduced in the step~1. 
In the following calculations, $B_b$ and $B_c$ denote, respectively, the bifurcation spheres of BEH and CEH, and $B_{cut}$ is the sphere of radius $r_{cut}$ which is the outer (inner) boundary of $\Sigma_b$ ($\Sigma_c$). 
As mentioned in Sec.\ref{sec:iw}, we use the Lagrangian $L(g,\Lambda)\,\vol$ in Eq.\eqref{eq:iw.L} with the variable $\Lambda$, which gives ${\bf E}_{\Lambda} = -(1/8\pi)\,\vol$.
The metric is given in Eq.\eqref{eq:mfl.metric}, and the timelike Killing vector $\partial_t$ is regarded as the symmetry generator $\xi \defeq \partial_t$.

Eq.\eqref{eq:iw.mfl_primitive} with $\Sigma_b$ in the SdS spacetime reads
\eqb
\label{eq:mfl.beh_primitive}
 -\frac{1}{8\pi}\int_{\Sigma_b} \xi\cdot\vol\,\delta\Lambda
  - \int_{B_b} \delta {\bf Q}_{\xi}
  + \int_{B_{cut}} \left[\, \delta {\bf Q}_{\xi}
                        - \xi\cdot\pot\left(g , \delta g \right) \, \right]  = 0 \, , 
\eqe
where $\xi = 0$ at $B_b$ is used in the second term, and the sign of the second and third terms are determined by the normal directions to $\Sigma_b$ at $B_b$ and $B_{cut}$ which are, respectively, in-ward and out-ward pointing along the ``$r$-axis''. 
The first term in the left-hand side of Eq.\eqref{eq:mfl.beh_primitive} is a simple three dimensional volume integral in $\Sigma_b$,
\eqb
\label{eq:mfl.beh_1}
 -\frac{1}{8\pi}\int_{\Sigma_b} \xi\cdot\vol\,\delta\Lambda
 =  -\frac{1}{6}\,r_{cut}^3 \,\delta\Lambda  + \frac{1}{6}\,r_b^3 \, \delta\Lambda .
\eqe

For the second term in the left-hand side of Eq.\eqref{eq:mfl.beh_primitive}, note that the symplectic potential ${\bf \Theta}$ includes the $(n-1)$-volume form as its factor by definition~\eqref{eq:iw.delta_L}, and consequently the Noether charge ${\bf Q}_{\xi}$ includes the $(n-2)$-volume form as its factor by definition~\eqref{eq:iw.Q}. 
($n=4$ for SdS spacetime.) 
This implies that the integral $\int$ and the variation $\delta$ of Noether charge is commutative, $\int_{B_b}\delta{\bf Q}_{\xi} = \delta\left(\int_{B_b}{\bf Q}_{\xi}\right)$. 
On the other hand, we can find for the Lagrangian in Eq.\eqref{eq:iw.L} with the variable~$\Lambda$ that the Noether charge ${\bf Q}_{\xi}$ takes the same form with the ordinary general relativity without $\Lambda$~\cite{ref:iw},
\eqb
\label{eq:mfl.Q}
 \left({\bf Q}_{\xi}\right)_{\mu\nu} = 
  -\frac{1}{16 \pi}\, \vol_{\mu\nu \alpha\beta} \nabla^{\alpha}\,\xi^{\beta} \, .
\eqe
Then, referring to the proof of the theorem~6.1 in the original Iyer-Wald formalism~\cite{ref:iw} which holds even with the variable $\Lambda$, we get
\eqb
 \delta\left[\int_{B_b} \left({\bf Q}_{\xi}\right)_{ab} \right]
  = - \frac{\kappa_b}{16 \pi} \,
      \delta\left( \int_{B_b} \,
      \binormal_{\alpha\beta}\,\left.\vol^{\alpha\beta}\right._{ab} \right) \, ,
\eqe
where the indices $a$ and $b$ are the \emph{abstract indices}~\cite{ref:Wald}, 
$\binormal_{\alpha\beta}$ is the bi-normal 2-form to $B_b$ which satisfies $\nabla_{\mu}\xi_{\nu} = \kappa_b\,\binormal_{\mu\nu}$ on $B_b$, and the vanishing variation of surface gravity at $B_b$ ($\delta\kappa_b\bigr|_{\rm B_b} = 0$) is used~\footnote{
By definition of $\kappa_b$ in Eq.\eref{eq:mfl.kappa_def} and $\delta\xi^{\mu} = 0$, we get $\xi^{\nu}\,(\delta\Gamma^{\mu}_{\nu\alpha})\,\xi^{\alpha} = (\delta\kappa_b)\,\xi^{\mu}$ at the BEH, where $\Gamma^{\mu}_{\nu\alpha}$ is the Christoffel symbol. 
When this is evaluated by, for example, the Kruskal-Szekeres type coordinate covering the inside and outside of BEH, we can find the limiting behavior $\delta\kappa_b \to 0$ as approaches the bifurcation sphere $B_b$ in BEH, i.e. $\delta\kappa_b\bigr|_{B_b} = 0$.
}. 
Hence we obtain for the second term in Eq.\eref{eq:mfl.beh_primitive},
\eqb
\label{eq:mfl.beh_2}
 \int_{B_b} \delta {\bf Q}_{\xi}
  = \frac{\kappa _b}{16 \pi}\delta\left( 2 \int_{B_b} d^2\Omega_b \right)
  = \frac{\kappa _b}{2 \pi}\delta\left( \pi r_b^2 \right) \, ,
\eqe
where $d^2\Omega_b \defeq r_b^2\,\sin\theta\,d\theta\,d\varphi$ is the area element on $B_b$.

Let us proceed to the calculation of the third term in the left-hand side of Eq.\eqref{eq:mfl.beh_primitive}.
For the first we calculate the integral of Noether charge $\int_{B_r} {\bf Q}_{\xi}$, where $B_r$ is a two-sphere of an arbitrary radius~$r$. 
Note that the $\theta$-$\varphi$ component of ${\bf Q}_{\xi}$ in Eq.\eqref{eq:mfl.Q} contributes to the integral,
\eqab
 \left({\bf Q}_{\xi}\right)_{\theta\varphi}
 = - \frac{1}{16 \pi}\, \vol_{\theta\varphi\,\alpha\beta}
     \nabla^{\alpha}\,\xi^{\beta} 
 = - \frac{1}{16 \pi}\, \pd{g_{tt}}{r}\,\xi^t\,
     \sqrt{-\det (g_{\mu\nu})}\,\,d\theta \wedge d\varphi \, .
\eqae
This gives
\eqb
\label{eq:mfl.int_Q}
 \int_{B_r} {\bf Q}_{\xi}
  = -\frac{1}{16 \pi} \int_{B_r} d^2\Omega_r\, \pd{g_{tt}}{r}\,\xi^t
  = \frac{1}{16 \pi} \,\frac{d f(r)}{dr}\, 4\pi r^2 \, ,
\eqe
where $d^2\Omega_r \defeq r^2\,\sin\theta\,d\theta\,d\varphi$ is the area element on $B_r$ and the definitions $\xi \defeq \partial_t$ and $g_{tt} = -f(r)$ in Eq.\eqref{eq:mfl.f} are used. 
Then, using the relation $\int_{B_r} \delta {\bf Q}_{\xi} = \delta\left(\int_{B_r}{\bf Q}_{\xi} \right)$, the first integral in the third term in Eq.\eqref{eq:mfl.beh_primitive} becomes
\eqb
 \int_{B_{cut}} \delta {\bf Q}_{\xi}
  = \delta \int_{B_{cut}} {\bf Q}_{\xi}
  = \delta\left( \frac{r_{cut}^2 f'(r_{cut})}{4} \right)
  = \frac{r_{cut}^2}{4}\, \delta f'(r_{cut}) \, ,
\eqe
where $f' \defeq df/dr$ and Eq.\eqref{eq:mfl.int_Q} is used in the second equality. 
Next, the explicit form of the symplectic potential $\pot$ for the Lagrangian~\eqref{eq:iw.L} with the variable $\Lambda$ is
\eqb
 \pot_{\lambda\mu\nu} =
   \frac{1}{16 \pi} \vol_{\lambda\mu\nu \alpha}\, g^{\alpha\beta} g^{\sigma\tau}\,
   \left[\, \nabla_{\tau} ( \delta g_{\beta\sigma} )
          - \nabla_{\beta} ( \delta g_{\sigma\tau} )\, \right]  \, .
\eqe
This is the same form with the ordinary general relativity without $\Lambda$~\cite{ref:iw}. 
Here, under the variations of $\delta M$ and $\delta \Lambda$, the metric variation $\delta g_{\mu\nu}$ is given from Eq.\eqref{eq:mfl.metric},
\eqb
\label{eq:mfl.dg}
 \delta(ds^2) = \delta g_{\mu\nu}\,dx^{\mu}\,dx^{\nu}
 = -\delta f(r)\,dt^2 + \frac{-\delta f(r)}{f^2(r)}\,dr^2 \, ,
\eqe
where the spherical part vanishes since that part does not depend on $M$ and $\Lambda$, and
\eqb
 \delta f(r) = - \frac{2}{r}\,\delta M - \frac{r^2}{3}\,\delta\Lambda \, .
\eqe
It should be emphasized here that the metric variation~\eref{eq:mfl.dg} is a static solution of the extended linearized Einstein equation~\eref{eq:iw.linear} and satisfies the condition to ensure Eq.\eref{eq:iw.mfl_primitive}. 
Furthermore note that, for the integrand $\xi^{\alpha}\,\pot_{\alpha\mu\nu}$ in the integral $\int_{B_{cut}} \xi\cdot\pot$, the indices $\mu$ and $\nu$ denotes the tangential components to $B_{cut}$ ($\theta$-$\varphi$ component), $\xi^{\alpha}\,\pot_{\alpha\,\theta\varphi}$. 
Then we obtain
\eqab
 \int_{B_{cut}} \xi\cdot\pot
 &=& -\frac{1}{16 \pi}\int_{B_{cut}} d^2\Omega_{cut} \, g^{r \beta} g^{\sigma\tau}\,
      \left[\, \nabla_{\tau} ( \delta g_{\beta\sigma} )
             - \nabla_{\beta} ( \delta g_{\sigma\tau} )\, \right] \nonumber \\
 &=& -\frac{1}{16 \pi}\int_{B_{cut}} d^2\Omega_{cut}
      \left[ -\delta f'(r) - \frac{2}{r}\,\delta f(r) \right] \nonumber \\
 &=& \frac{r_{cut}^2}{4}\,
     \left[ \delta f'(r_{cut}) + \frac{2}{r_{cut}}\,\delta f(r_{cut}) \right] \, ,
\eqae
where $d^2\Omega_{cut} \defeq r_{cut}^2\,\sin\theta\,d\theta\,d\varphi$ is the area element on $B_{cut}$, and $\xi \defeq \partial_t$ and $\vol_{t \mu\nu r} = - \vol_{tr \mu\nu}$ are used in the first equality. 
Hence the third term in Eq.\eqref{eq:mfl.beh_primitive} becomes
\eqb
\label{eq:mfl.beh_3}
 \int_{B_{cut}}
 \left[\, \delta {\bf Q}_{\xi}
       - \xi\cdot\pot\left(g , \delta g \right) \,\right]
 = \delta M + \frac{1}{6}\,r_{cut}^3\,\delta\Lambda \, .
\eqe

Then, collecting Eqs.\eqref{eq:mfl.beh_1}, \eqref{eq:mfl.beh_2} and \eqref{eq:mfl.beh_3}, we obtain from Eq.\eqref{eq:mfl.beh_primitive},
\eqb
\label{eq:mfl.beh}
 \delta M = \frac{\kappa_b}{2 \pi}\,\delta\left( \pi r_b^2\right)
            - \frac{1}{6}\, r_b^3 \, \delta\Lambda \, .
\eqe
It should be noted that the working variable $r_{cut}$ does not appear in this equation.

Next turn our calculation to Eq.\eqref{eq:iw.mfl_primitive} with the hypersurface $\Sigma_c$, which reads 
\eqb
\label{eq:mfl.ceh_primitive}
 -\frac{1}{8\pi}\int_{\Sigma_c} \xi\cdot\vol\,\delta\Lambda
  + \int_{B_c} \delta {\bf Q}_{\xi}
  - \int_{B_{cut}} \left[\, \delta {\bf Q}_{\xi}
                        - \xi\cdot\pot\left(g , \delta g \right) \, \right]  = 0 \, , 
\eqe
where $\xi = 0$ at $B_c$ is used in the second term, and the sign of the second and third terms are determined by the normal directions to $\Sigma_c$ at $B_c$ and $B_{cut}$ which are, respectively, out-ward and in-ward pointing along the ``$r$-axis''. 
Then, following the same calculations to obtain Eq.\eqref{eq:mfl.beh}, we obtain from Eq.\eqref{eq:mfl.ceh_primitive},
\eqb
\label{eq:mfl.ceh}
 \delta M = \frac{\kappa_c}{2 \pi}\,\delta\left( \pi r_c^2\right)
            - \frac{1}{6}\, r_c^3 \, \delta\Lambda \, .
\eqe
It should be noted that the working variable $r_{cut}$ does not appear in this equation.

So far we have carried out the step~2 to obtain Eqs.\eqref{eq:mfl.beh} and~\eqref{eq:mfl.ceh}. 
Before proceeding to the step~3 of our strategy, let us comment on Ref.\cite{ref:sds.mfl_2}: 
As mentioned in Sec.\ref{sec:intro}, Ref.\cite{ref:sds.mfl_2} has already discussed the MFL of SdS spacetime with variable $\Lambda$ using some integral quantities as state variables, and obtained two MFLs separately for BEH and CEH. 
Those MFLs in~\cite{ref:sds.mfl_2} are the same with our Eqs.\eqref{eq:mfl.beh} and~\eqref{eq:mfl.ceh}. 
Although the derivation of MFLs in~\cite{ref:sds.mfl_2} includes the problem of the choice of integration constant, but our derivation based on the extended Iyer-Wald formalism is free from such problem. 
Furthermore, the integral quantity used in~\cite{ref:sds.mfl_2} requires to place a boundary (corresponding to $B_{cut}$ in our step~2) at CEH in deriving Eq.\eqref{eq:mfl.beh} and at BEH in deriving Eq.\eqref{eq:mfl.ceh}. 
If we interpret the boundary as the position of an observer, it seems not to be physically acceptable to place the observer at the event horizon. 
However our surface $B_{cut}$ may be appropriate as a candidate of the position of observer who measures thermodynamical quantities of two-horizon system.

Although Eqs.\eqref{eq:mfl.beh} and~\eqref{eq:mfl.ceh} are regarded as MFLs of SdS spacetime in Ref.\cite{ref:sds.mfl_2}, our aim is to propose a single formula~\eqref{eq:mfl.mfl_def} as the appropriately expressed MFL of SdS spacetime. 
The procedure to obtain the appropriate MFL is the step~3 of our strategy. 
Then, let us carry out the step~3: 
Combining Eqs.\eqref{eq:mfl.beh} and~\eqref{eq:mfl.ceh} together with the definition~\eqref{eq:mfl.S}, we get
\eqb
\label{eq:mfl.dS}
 \delta S
 = 2 \pi \left( \frac{1}{\kappa_b} + \frac{1}{\kappa_c} \right) \,\delta M
  + \frac{\pi}{3} \left( \frac{r_b^3}{\kappa_b} + \frac{r_c^3}{\kappa_c} \right) \,
    \delta\Lambda \, .
\eqe
On the other hand, we get from Eq.\eqref{eq:mfl.M.Lambda},
\eqb
 \delta r_b = \frac{2\,\delta M + (r_b^3/3)\,\delta\Lambda}{1 - \Lambda\,r_b^2}
 \quad,\quad
 \delta r_c = \frac{2\,\delta M + (r_c^3/3)\,\delta\Lambda}{1 - \Lambda\,r_c^2} \, .
\eqe
These variations of horizon radii together with the definition~\eqref{eq:mfl.V} give the volume variation,
\eqb
\label{eq:mfl.delta_V}
 \delta V
 = 4 \pi \left( \frac{r_c}{\kappa_c} - \frac{r_b}{\kappa_b} \right)\,\delta M
  + \frac{2 \pi}{3}
    \left( \frac{r_c^4}{\kappa_c} - \frac{r_b^4}{\kappa_b} \right)\,\delta\Lambda \, .
\eqe
Then, substituting this $\delta V$ into Eq.\eqref{eq:mfl.dS}, we obtain the appropriate MFL of SdS spacetime~\eqref{eq:mfl.mfl_def},
\eqb
\label{eq:mfl.mfl}
 \delta M = \Teff\,\delta S - \peff\,\delta V \, ,
\eqe
where the coefficients are
\eqab
\label{eq:mfl.Teff}
 \Teff &=&
   \frac{r_b^4}{(r_c + r_b)\,(r_c^3 - r_b^3)} \,\frac{|\kappa_c|}{2\,\pi}
    + \frac{r_c^4}{(r_c + r_b)\,(r_c^3 - r_b^3)} \,\frac{\kappa_b}{2\,\pi} \nonumber \\
 &=&
   \frac{1}{4 \pi \,r_c}\,\frac{x^4 + x^3 -2 x^2 + x + 1}{x\, (x+1)\, (x^2 + x + 1)} \, ,\\
\label{eq:mfl.peff}
 \peff &=&
  \frac{1}{8 \pi \, r_c^2}\,
  \frac{(1-x)\,(x^4 + 3 x^3 + 3 x^2 + 3 x + 1)}{x\,(x+1)\,(x^2 + x + 1)^2} \, ,
\eqae
where $x \defeq r_b/r_c$ and $0 < x < 1$. 
We find that $\Teff$ and $\peff$ are positive definite, $\Teff > 0$ and $\peff > 0$, and the requirement~\eqref{eq:mfl.req} is satisfied. 
Hence the MFL in Eq.\eqref{eq:mfl.mfl} with the coefficients given above is the appropriately expressed MFL of SdS spacetime.

Here note that, although we refer to the entropy-area law of asymptotic flat black hole thermodynamics as a motivation to adopt the definition~\eqref{eq:mfl.S}, it does not mean to assume~$S$ in Eq.\eqref{eq:mfl.S} to be the physical entropy of SdS spacetime. 
It is a future task to resolve the issue whether this $S$ is really a physical entropy in SdS thermodynamics. 
Even if $S$ is not a physical entropy, our MFL in Eq.\eqref{eq:mfl.mfl} suggests the \emph{mass formula} which relates mass parameter $M$ and total horizon area $S$. (The mass formula of asymptotic flat black holes is the Smarr's formula~\cite{ref:mf.smarr}.)

Our derivation of the appropriately expressed MFL~\eqref{eq:mfl.mfl} is based on the extended Iyer-Wald formalism. 
However, since the relation among $M$, $S$ and $V$ is definitely determined by the metric~\eqref{eq:mfl.metric}, the relation~\eqref{eq:mfl.mfl} must be obtained from $S$ and $V$ defined in Eqs.~\eqref{eq:mfl.S} and~\eqref{eq:mfl.V} without using the extended Iyer-Wald formalism. 
Indeed, Ref.\cite{ref:sds.mfl_2} has already obtained Eqs.\eqref{eq:mfl.beh} and~\eqref{eq:mfl.ceh} (whose combination gives Eq.\eqref{eq:mfl.mfl}) by using some integral quantities as state variables. 
The advantages of using the extended Iyer-Wald formalism are the following three points: (1)~it is free from the problem of integration constant as mentioned in Sec.\ref{sec:intro}, (2)~it offers naturally the definition of $V$ in Eq.\eqref{eq:mfl.V}, and (3)~it gives the positive definite effective temperature $\Teff$ and pressure $\peff$ which, together with $V$, are suitable to describe the SdS black hole evaporation process as discussed in the next section.

Finally in this section, for the completeness of our discussion, let us consider the case that $\Lambda$ is not regarded as an independent variable. 
In this case, one obtains a relation, $\delta S = 2\pi\,(1/\kappa_b + 1/\kappa_c)\,\delta M$, from the definition~\eqref{eq:mfl.S}. 
By rearranging this relation appropriately and introducing $V$ defined in Eq.\eqref{eq:mfl.V}, one can formally obtain the MFL in Eq.\eqref{eq:mfl.mfl} with the same coefficients in Eqs.\eqref{eq:mfl.Teff} and~\eqref{eq:mfl.peff}. 
However in this case, we find relations; $\partial M(S,V)/\partial S = (dS/dM)^{-1} \not\equiv \Teff$ and $- \partial M/\partial V = -(dV/dM)^{-1} \not\equiv -\peff$. 
These contradict the definition of coefficients $\Teff$ and $\peff$ in Eq.\eqref{eq:mfl.mfl}. 
By the reductive absurdity, this fact indicates the necessity of ``variable $\Lambda$'' for a mathematically consistent MFL of SdS spacetime of the form in Eq.\eqref{eq:mfl.mfl}. 
Here note that the same claim, the necessity of variable $\Lambda$, is already suggested in~\cite{ref:sds.mfl_2}, although the MFL in~\cite{ref:sds.mfl_2} is given in a different expression from Eq.\eqref{eq:mfl.mfl} and includes the uncertainty of integration constant of conserved quantities as mentioned in Sec.\ref{sec:intro}. 
The necessity of variable $\Lambda$ seems a universal fact for mathematically consistent MFL of SdS spacetime.

\section{Discussions}
\label{sec:d}

Let us discuss what is implied by our appropriate MFL in Eq.\eqref{eq:mfl.mfl} if it is regarded as a thermodynamical first law of SdS spacetime. 
As an interesting process, we treat the SdS black hole evaporation process at constant $\Lambda$.

Before considering SdS black hole evaporation, let us recall the so-called \emph{generalized second law} in the evaporation process of Schwarzschild black hole. 
In Schwarzschild thermodynamics, the thermodynamical first law is given by Eq.\eqref{eq:mfl.mfl_flat} with setting the ``work terms'' zero, $\delta M = T_{\rm BH}\,\delta S_{\rm BH}$, where $T_{\rm BH} \defeq \kappa_{\rm BH}/2 \pi = 1/8 \pi M$ is the Hawking temperature and $S_{\rm BH} \defeq A_{\rm BH}/4 = \pi (2 M)^2$ is the black hole entropy. 
When Schwarzschild black hole evaporates, the mass energy $M$ decreases and consequently $S_{\rm BH}$ decreases. 
Here note that, the evaporation of isolated black hole is an irreversible adiabatic process. 
Hence, if the total entropy of the whole system is given by $S_{\rm BH}$, the decrease of $S_{\rm BH}$ contradicts the second law of thermodynamics which requires the increase of total entropy for irreversible adiabatic processes. 
Then the generalized second law claims that the total entropy of black hole and matter fields of Hawking radiation increases for the evaporation process of isolated black holes~\cite{ref:gsl,ref:gsl_2}.
Therefore, at least for the evaporation of asymptotic flat black holes, the generalized second law is necessary to hold the validity of thermodynamical formulation of black holes.

Then, proceed to the discussion of SdS black hole evaporation at constant $\Lambda$. 
When SdS black hole evaporates, it seems reasonable to consider the mass parameter decreases, $\delta M < 0$. 
Here, for any process at constant~$\Lambda$, our MFL in Eq.\eqref{eq:mfl.mfl} and $\delta V$ in Eq.\eqref{eq:mfl.delta_V} are rearranged to 
$\delta S = 2 \pi\,( 1/\kappa_b + 1/\kappa_c ) \,\delta M$ and $\delta V = 4 \pi\,( r_c/\kappa_c - r_b/\kappa_b ) \,\delta M$. 
Then, we get $\delta S > 0$ and $\delta V > 0$ due to $\delta M < 0$ and Eq.\eqref{eq:mfl.kappa_relations}. 
The expansion of volume $\delta V > 0$ is a reasonable result if $\peff$ is interpreted as a pressure. 
The increase of entropy $\delta S > 0$ denotes that the generalized second law is not needed to ensure the second law of SdS thermodynamics for its evaporation process at constant~$\Lambda$. 
Hence our appropriate MFL in Eq.\eqref{eq:mfl.mfl} suggests a thermodynamically consistent description of SdS black hole evaporation at constant $\Lambda$.

Finally let us comment on the coefficient $\Teff$ in Eq.\eqref{eq:mfl.Teff}. 
This $\Teff$ is regarded as a temperature if our MFL in Eq.\eqref{eq:mfl.mfl} is regarded as a thermodynamical first law. 
Then one may wonder about what the physical meaning of the temperature is, since the SdS spacetime is essentially a non-equilibrium system due to the difference of Hawking temperatures of BEH and CEH~\cite{ref:sds.temperatures}. 
Here let us point out that there is a long history of research on two-temperature non-equilibrium systems, and there are many proposals on the definition of non-equilibrium temperature of non-equilibrium systems (see, for example, references of works in~\cite{ref:gsl_2}). 
No commonly accepted definition of non-equilibrium temperature exists as present. 
Therefore, from the point of view of non-equilibrium thermodynamics, the suggestion of defining a non-equilibrium temperature is meaningful at present. 
If the SdS thermodynamics is formulated with regarding our MFL in Eq.\eqref{eq:mfl.mfl} as a thermodynamical first law, then our $\Teff$ in Eq.\eqref{eq:mfl.Teff}, which can be expressed as a linear combination of the surface gravities of BEH and CEH, may be understood as one suggestion of an effective temperature of a non-equilibrium system. 
At present, the physical meaning of $\Teff$ is an open issue which requires further researches on non-equilibrium nature of SdS spacetime. 
The research on SdS thermodynamics provides the stage for the intersection of gravitational physics and non-equilibrium physics.

\ack

One of authors H.S. is supported by the Grant-in-Aid for Scientific Research Fund of the Ministry of Education, Culture, Sports, Science and Technology, Japan (Young Scientists (B) 19740149).


\end{document}